\documentclass[12pt,a4paper]{article}

\usepackage{amsmath,amssymb}

\begin{document}

\bibliographystyle{unsrt}

\title{Integrable systems on the lattice and orthogonal polynomials of discrete variable}

\author{Miguel Lorente\\
\small Departamento de F\'{\i}sica, Universidad de Oviedo, 33007 Oviedo, Spain}
\date{}
\maketitle

\begin{abstract}
Some particular examples of classical and quantum systems on the lattice are solved with the help of
orthogonal polynomials and its connection to continuous models are explored.
\end{abstract}


\section{Introduction}

Recently there is an increasing interest on discrete models in classical and quantum physics. Numerical
calculations use computational technics to solve differential equations [2]. Lattice field theories have
become a very powerfull tool to avoid infinities in perturbative methods, and to obtain exact solutions of
the field equations [7]. Ponzano-Regge calculus introduced in gravitational field is equivalent to Penrose
spin network that discretizes riemannian manifolds [1]. Statistical mechanics has been working from the
begining with lattice approximation. 

The orthogonal polynomials of discrete variable [8] offers a new approach to these models. They are exact
solutions of difference equations from which raising and lowering operators, eigenvalue equation,
symmetries and constant of motion can be calculated. If a physical problem is not given in a discrete
form, this can be guessed if we know the differential equation of some orthogonal polynomial which is the
continuous limit of the diference equation.

We present some simple examples of discrete models in classical and quantum systems, that can be solved by
the above method. In the last case, special care has to be taken in the construction of Hilbert space when
the basis are the orthogonal polynomials of discrete variables.

\section{Orthogonal polynomials of discrete variable}

A polynomial of hypergeometric type $P_n (x)$ of discrete variable $x$ satisfies two
fundamental relations from which one derives raising and lowering operators [6]:

\medskip
\noindent i)  Difference equation:
\begin{equation}
\sigma (x)\Delta \nabla P_n (x) + \tau (x)\Delta P_n (x) + \lambda _n P_n
(x) = 0
\end{equation}
where $\sigma (x)$ and $\tau (x)$ are polynomials of, at most, second and first degree,
respectively, $\Delta (\nabla)$ are the forward (backward) operators and $\lambda _n$ is the eigenvalue corresponding to the
eigenfunction $P_n (x)$.

\medskip
\noindent ii)   Three term recurrence relations:
\begin{equation}
xP_n (x) = \alpha _n P_{n + 1} (x) + \beta _n P_n (x) + \gamma _n P_{n
- 1} (x)
\end{equation}

\medskip
\noindent iii)    Raising operator
\smallskip

\begin{equation}
\sigma (x)\nabla P_n (x) = \frac{{\lambda _n }}{{n\tau '_n }}\tau _n (x)P_n
(x) -
\frac{{\lambda _{2n} }}{{2n}}P_{n + 1} (x)
\end{equation}
where
\begin{eqnarray*}
 \tau _n (x) &=& \tau (x + n) + \sigma (x + n) - \sigma (x) 
\end{eqnarray*}

\medskip
\noindent iv) Lowering operator
\smallskip

\begin{align}
&\left( {\sigma (x) + \tau (x)} \right)\Delta P_n (x) = \left[ { -
\frac{{\lambda _n }}{n}\frac{{2n + 1}}{{\lambda_{2n + 1} }}\tau (x) - \lambda _n  -
\frac{{\lambda _{2n} }}{{2n}}(x - \beta _n )} \right]P_n (x) \nonumber \\
&\qquad + \frac{{\lambda _{2n}
}}{{2n}}\gamma _n P_{n - 1} (x)
\end{align}

These polynomials are orthogonal with respect to the weight function $\rho (x)$, particular examples of which are the Kravchuk,
Meixner, Charlier, and Hahn polynomials.

From the orthogonal polynomials of discrete variable we can construct the corresponding orthonormal functions of discrete variable, by
the definition
\begin{equation}
\phi _n (x) = d_n^{ - 1} \sqrt {\rho (x)} P_n (x)
\end{equation}
where $d_n$ is some normalization constant. They satisfy

\medskip
\noindent i)  Difference equation
\begin{align}
\hspace*{-10mm} &\sqrt {\left( {\sigma (x) + \tau (x)} \right)\sigma (x + 1)} \phi _n
(x + 1) +\sqrt {\left( {\sigma (x - 1) + \tau (x - 1)} \right)\sigma (x)} \phi _n (x -
1) \nonumber \\
& \qquad - \left( {2\sigma (x) + \tau (x)} \right)\phi _n (x) + \lambda _n \phi _n (x) = 0
\end{align}

\noindent ii)  Three term recurrence relation:
\begin{eqnarray}
\frac{{\lambda _{2n} }}{{2n}}\alpha _n \frac{{d_{n + 1} }}{{d_n }}\phi
_{n + 1} (x) + \frac{{\lambda _{2n} }}{{2n}}\gamma _n \frac{{d_{n - 1} }}{{d_n }}\phi _{n
- 1} (x) + \frac{{\lambda _{2n} }}{{2n}}(\beta _n  - x)\phi _n (x) = 0
\end{eqnarray}

\noindent iii)  Raising operator
\begin{align}
\hspace*{-10mm} &L^ +  (x,n) \equiv  \left[ {\frac{{\lambda _n }}{n}\frac{{\tau _n
(x)}}{{\tau '_n }} - \sigma (x)} \right]\phi _n (x) + \sqrt {\left( {\sigma (x - 1) + \tau
(x - 1)} \right)\sigma (x)} \phi _n (x - 1) \nonumber  \\
&\qquad  = \frac{{\lambda _{2n} }}{{2n}}\alpha _n
\frac{{d_{n + 1} }}{{d_n }}\phi _{n + 1} (x)
\end{align}

\noindent iv)  Lowering operator
\begin{eqnarray}
L^ -  (x,n)\equiv \left[ { - \frac{{\lambda _n }}{n}\frac{{\tau _n
(x)}}{{\tau '_n }} + \lambda _n  + \frac{{\lambda _{2n} }}{{2n}}(x - \beta _n ) - \sigma
(x) - \tau (x)} \right]\phi _n (x)   \nonumber \\ 
  \qquad + \sqrt {\left( {\sigma (x) + \tau (x)} \right)\sigma (x + 1)} \phi _n (x + 1) =
\frac{{\lambda _{2n} }}{{2n}}\gamma _n \frac{{d_{n - 1} }}{{d_n }}\phi _{n - 1} (x) 
\end{eqnarray}

The difference equation i) written in the form  $H(x,n)\phi_n(x)=0$, can be factorized with the help of the raising and lowering
operators:
\begin{eqnarray}
L^ -  (x,n+1)L^ +  (x,n)=\mu (n)+u(x+1,n)H(x,n) \\[1ex]
L^ +  (x,n)L^ -  (x,n+1)=\mu (n)+u(x,n-1)H(x,n+1)
\end{eqnarray}
where 
$\mu (n)=\displaystyle {{\lambda _{2n}} \over {2n}}{{\lambda _{2n+2}} \over {2n+2}}\ \alpha_n\gamma
_{n+1},\quad u(x,n)={{\lambda _n\tau _n(x)} \over {n\tau '_n}}-\sigma (x)$

\section{The quantum harmonic oscillator of discrete \\variable}

We start from the orthogonal polynomials of a discrete variable, the Kravchuk
polynomials $k_n^{(p)} (x)$ and the corresponding orthonormal Kravchuk functions [5]
\begin{equation}
K_n^{(p)} (x) = d_n^{ - 1} \sqrt {\rho (x)\;} k_n^{(p)} (x),
\end{equation}
where $d_n^2  = \frac{{N!}}{{n!(N - n)!}}(pq)^n $ is a normalization constant, $\rho
(x) = \frac{{N!p^x q^{N - x} }}{{x!(N - x)!}}(pq)^n $ is the weight function, with $p >
0,\quad q > 0,\quad p + q = 1,\quad x = 0,1, \ldots N - 1.$

From the difference equation
\begin{align}
 \hspace{-10mm}&\sqrt {pq(N - x)(x + 1)} K_n^{(p)} (x + 1) \nonumber  \\ 
&\qquad +  \sqrt {pq(N - x + 1)x} K_n^{(p)} (x
- 1)  + \left[ {x(p - q) - Np + n} \right] K_n^{(p)} (x) = 0,
\end{align}
and the recurrence relation
\begin{align}
 \hspace{-10mm}&\sqrt {pq(N - n)(x + 1)} K_{n + 1}^{(p)} (x) \nonumber  \\ 
 &\qquad + \sqrt {pq(N - n + 1)n} K_{n -1}^{(p)} (x) + \left[ {n(q - p) + Np - x} \right]K_n^{(p)} (x) = 0,
\end{align}
we construct raising and lowering operators
\begin{align}
& L^ +  (x,n)K_n^{(p)} (x) = pq(x + n - N)K_n^{(p)} (x)  \nonumber \\ 
&\qquad + \sqrt {pq(N - x + 1)x} K_n^{(p)} (x - 1) = \sqrt {pq(N - n)(n + 1)} K_{n + 1}^{(p)} (x),\\[1ex]
& L^ -  (x,n)K_n^{(p)} (x) = pq(x + n - N)K_n^{(p)} (x) \nonumber\\ 
&\qquad + \sqrt {pq(N - x)(x + 1)}
K_n^{(p)} (x + 1)   = \sqrt {pq(N - n + 1)n} K_{n - 1}^{(p)} (x).
\end{align}
As in the general case, these operators factorize the difference equation
\begin{eqnarray}
 \hspace{-10mm} L^ +  (x,n - 1)L^ -  (x,n) = pq(N - n + 1)n + pq(x + n - 1 - N)H(x,n), \\[1ex]
 \hspace{-10mm} L^ -  (x,n + 1)L^ +  (x,n) = pq(N - n)(n + 1) + pq(x + n + 1 - N)H(x,n).
\end{eqnarray}

In order to justify the name of quantum oscillator of discrete variable we substitute 
$x=Np+\sqrt {2Npq}s$, and take the limit $N\to \infty $ in the former expressions. We get
\begin{equation}
K_n^{(p)}(x)\longrightarrow \sqrt {{1 \over {2^nn!\pi }}}e^{-{{s^2} \mathord{\left/ {\vphantom
{{s^2} 2}}
\right. \kern-\nulldelimiterspace} 2}}H_n(s)\equiv \psi (s)
\end{equation}
 where $H_n(s)$ is the Hermite polynomial of continuous variable. The difference and
recurrence relations for the Kravchuk functions becomes in the limit the differential and
recurrence relations for the normalized Hermite functions $\psi (s)$ which are the solution
of the quantum harmonic oscillator.

In the limit the raising and lowering operators becomes the creation and annihilation
operators
\begin{equation}
{1 \over {\sqrt {Npq}}}\;L^+(x,n)\;K_n^{(p)}(x)\mathrel{\mathop{\kern0pt\longrightarrow}\limits_{N\to \infty }}{1 \over 2}\left\{ {s-{d \over
{ds}}} \right\}\psi _n(s)\equiv a^+\psi (s)
\end{equation}
\begin{equation}
{1 \over {\sqrt {Npq}}}\;L^-(x,n)\;K_n^{(p)}(x)\mathrel{\mathop{\kern0pt\longrightarrow}\limits_{N\to \infty }}{1 \over 2}\left\{ {s+{d \over
{ds}}} \right\}\psi _n(s)\equiv a\psi (s)
\end{equation}
The commutator of the raising and lowering operators are closed under the $SO(3)$ algebra.
\begin{equation}
{1 \over {Npq}}\left[ {L,L^+} \right]K_n^{(p)}(x)=\left( {1-{n \over j}} \right)K_n^{(p)}(x)\equiv L^3K_n^{(p)}(x)
\end{equation}
which in the limit becomes $\left[ {a,a^+} \right]\psi (s)=\psi (s)$

Similarly the anticommutation relation of the raising and lowering operators becomes in the limit the hamiltonian of the quantum oscillator
\begin{align}
&{1 \over {Npq}}\left[ {L,L^+} \right]K_n^{(p)}(x)={1 \over j}\left( {j(j+1)-(j-n)^2} \right)K_n^{(p)}(x)\nonumber \\
&\qquad \longrightarrow
\left( {aa^++a^+a}
\right)\psi _n(s)=(2n+1)\psi _n(s)
\end{align}

\section{The hidrogen atom of discrete variable}

We define the orthonormal Meixner functions [5]
\begin{equation}
M_n^{(\gamma ,\mu )} (x) \equiv d_n^{ - 1} \sqrt {\rho _1 (x)} \;m_n^{(\gamma
,\mu )} (x)
\end{equation}
where $m_n^{(\gamma ,\mu )} (x)$
 are the Meixner polynomials, 
\[ d_n  = \frac{{n!\Gamma (n + \gamma )}}{{\mu ^n (1 - \mu )^\gamma  \Gamma (\gamma
)}},\quad \rho _1 (x) = \frac{{\mu ^x \Gamma (x + \gamma  + 1)}}{{\Gamma (x +
1)\Gamma (\gamma )}},
\]
and $\gamma ,\mu$ are real constants $0 < \mu  < 1,\quad \gamma  > 0$. 

They satisfy the orthogonality condition
\[
\sum {M_n^{(\gamma ,\mu )} (x)\;} M_{n'}^{(\gamma ,\mu )}
(x)\frac{1}{{\mu (x + \gamma )}} = \delta _{nn'} 
\]
and the following properties: 

\medskip
\noindent i) Difference equation
\begin{align}
&\lefteqn{ \sqrt {\frac{{\mu (x + \gamma )(x + 1)(x + \gamma )}}{{x + \gamma  + 1}}} M_n (x + 1)}\nonumber  \\ 
  &\qquad  + \sqrt {\mu (x + \gamma )x} M_n (x - 1) - \left[ {\mu (x + \gamma ) + x - n(1 - \mu )} \right]M_n (x)=0
\end{align}
ii) Recurrence relation
\begin{align}
&- \sqrt {{\mu (n + \gamma )(n + 1)}} M_{n + 1} (x)  - \sqrt{\mu (n + \gamma  - 1)n} M_{n - 1} (x)+ \nonumber \\
 & \qquad + \left( {\mu x + \mu n + \mu \gamma  + n - x} \right)M_n (x) = 0 
\end{align}
iii) Raising operator
\begin{align}
 &\hspace{-10mm}L^ +  (x,n)\;M_n (x) =  - \mu (x + \gamma  + n)\;M_n (x) + \sqrt {\mu (x + \gamma )x}
\;M_n (x - 1)  \nonumber\\[0.5ex] 
  &\qquad  = \sqrt {\mu (n + \gamma )(n + 1)} \;M_{n + 1} (x) 
\end{align}
 iv) Lowering operator
\begin{align}
& L^ -  (x,n)\;M_n (x) =  - \mu (x + \gamma  + n)\;M_n (x)+ \nonumber\\
&\qquad + \sqrt{\frac{{\mu (x + \gamma )(x + 1)(x + \gamma ) }}{{x + \gamma  + 1}}}\;M_n (x
+ 1)  =  - \sqrt {\mu (n + \gamma  - 1)n} \;M_{n - 1} (x) 
\end{align}
Notice that we have omitted, for the sake of brevity, the superindices $(\gamma, \mu)$ in $M_n(x)$.

In order to make connection between the Meixner functions of discrete variable and Laguerre functions of continuous variable we substitute
$\gamma=\alpha+1$, $\mu=1-h$, $x=s/h$ in the former and take the limit $h\to 0$, $x\to \infty$, $hx \to s$;
\begin{equation}
M_n^{(\gamma ,\mu )}(x)\longrightarrow \sqrt {{{n!} \over {\Gamma (n+\alpha +1)}}e^{-s}s^{\alpha +1}}L_n^\alpha (s)\equiv \psi _n^\alpha (s)
\end{equation}
We take also the limit of the following expressions:

\medskip
\noindent i) Differential equation
\begin{equation}
\psi _n^{\alpha ''} (s) + \left[ {\frac{\nu }{s} - \frac{1}{4} - \frac{{\alpha ^2 
- 1}}{{s^2 }}} \right]\;\psi _n^\alpha  (s) = 0, \quad \nu=n+\frac{1}{2}(\alpha+1).
\end{equation}
ii) Recurrence relations
\begin{eqnarray}
 \lefteqn{- \sqrt {\left( {n + \alpha  + 1} \right)(n + 1)} \psi _{n + 1}^\alpha  (s)-} \nonumber \\
& & \qquad - \sqrt
{\left( {n + \alpha } \right)n} \psi _{n - 1}^\alpha  (s) + (2n + \alpha  + 1 -
s)\;\psi _n^\alpha  (s) = 0.
\end{eqnarray}
iii) Raising operator
\begin{eqnarray}
 L^ +  (s,n)\;\psi _n^\alpha  (s) &=&  - \frac{1}{2}(2n + \alpha  + 1 - s)\;\psi
_n^\alpha  (s) - s\frac{d}{{ds}}\;\psi _n^\alpha  (s) \nonumber \\ 
 & =&  - \sqrt {\left( {n + 1} \right)(n + \alpha  + 1)} \;\psi _{n + 1}^\alpha  (s) 
 \end{eqnarray}
iv) Lowering operator
\begin{eqnarray}
 L^ -  (s,n)\;\psi _n^\alpha  (s) &=&  - \frac{1}{2}(2n + \alpha  + 1 - s)\;\psi
_n^\alpha  (s) + s\frac{d}{{ds}}\;\psi _n^\alpha  (s) =  \nonumber\\ 
  &=&  - \sqrt {n(n + \alpha )} \;\psi _{n - 1}^\alpha  (s)
 \end{eqnarray}
If we substitute in the differential equation (30) $\alpha=2l+1$, $\nu=n+l+1$, we obtain the reduced radial equation for the hydrogen atom,
\begin{equation}
{{d^2u} \over {ds^2}}+\left[ {{\nu  \over s}-{1 \over 4}-{{l(l+1)} \over {s^2}}} \right]u(s)=0
\end{equation}
the solutions of which are given by the generalized Laguerre functions
\begin{equation}
u_{\nu l}(s)=\left\{ {{{(\nu -l-1)!} \over {(\nu +l)!}}} \right\}^{{1 \over 2}}s^{l+1}e^{{s \over 2}}L_{\nu -l-1}^{2l+1}(s)
\end{equation}
This correspondence shows that we can use the difference equation of Meixner function as quantum model of hydrogen atom of discrete variable

\section{Calogero-Sutherland model on the lattice}

We start with the difference equation for the Hahn polynomials of discrete variable, in the particulr case $\alpha=\beta=\lambda-{1 \over 2}$, namely, 
\begin{align}
&\left[ {x\left( {N-x-\lambda -{3 \over 2}} \right)+\left( {\lambda +{1 \over 2}} \right)\left( {N-1} \right)} \right]h_n^{\left(
{\lambda -{1 \over 2},\lambda -{1 \over 2}} \right)}(x+1)+ \nonumber \\
 &\qquad +x\left( {N+\lambda -{1 \over 2}-x} \right) h_n^{\left( {\lambda -{1 \over 2},\lambda -{1
\over 2}} \right)}(x-1)- \nonumber\\
&\qquad -\left[ {2x\left( {N-\lambda -1} \right)+\left( {\lambda +{1 \over 2}} \right)\left( {N-1} \right)}
\right]h_n^{\left( {\lambda -{1
\over 2},\lambda -{1 \over 2}} \right)}(x)+ \nonumber\\
&\qquad + n(n+2\lambda )h_n^{\left( {\lambda -{1 \over 2},\lambda -{1 \over 2}} \right)}(x)=0
\end{align}
In the continuous limit $N \to \infty, {x \over N} \to s$, the Hahn polynomials become the Jacobi polynomials of continuous variable, that in the
particular case $\alpha=\beta=\lambda-{1 \over 2}$ are proportional to the Gegenbauer polynomials, namely,
\begin{equation}
h_n^{\left( {\lambda -{1 \over 2},\lambda -{1 \over 2}} \right)}(x)\longrightarrow P_n^{\left( {\lambda -{1 \over 2},\lambda -{1 \over 2}}
\right)}(s)={{\left( {\lambda +{1 \over 2}} \right)_n} \over {(2\lambda )_n}}C_n^\lambda (s)
\end{equation}
The difference equation for the Hahn polynomials becomes in the continuous limit the differential equation for the Gegenbauer polynomials
\begin{equation}
\left( {s^2-1} \right){{d^2} \over {ds^2}}C_n^\lambda (s)+\left( {2\lambda +1} \right)s{d \over {ds}}C_n^\lambda (s)=n\left( {n+2\lambda }
\right)C_n^\lambda (s)
\end{equation}
Using polar coordinates $s=\cos q$, this equation becomes, 

\begin{equation*}
-{{d^2} \over {dq^2}}C_n^\lambda (q)-2\lambda \cot q{d \over {dq}}C_n^\lambda (q)= \epsilon _n(\lambda )C_n^\lambda (q)
\end{equation*}
with $\epsilon_n (\lambda )=E_n(\lambda )-E_0=\left( {n+\lambda } \right)^2-\lambda ^2=n(n+2\lambda)$

If we normalize the solution by the weight function $\rho (q)=(\sin q)^\lambda $, that is, 
\begin{equation}
\psi _n^\lambda (q)=d_n(\sin q)^\lambda C_n^\lambda (q)
\end{equation}
we get the standard differential equation for the Calogero-Sutherland model [9] in one dimension
\begin{equation*}
H\psi _n^\lambda (q)=E_n(\lambda )\psi _n^\lambda (q),\qquad E_n(\lambda )=\left( {n+\lambda } \right)^n
\end{equation*}
\begin{equation}
H=-{{d^2} \over {dq^2}}-\lambda (\lambda -1){1 \over {\sin ^2q}}
\end{equation}

\section{Discrete time quantum mechanical systems}

Let $H(q,p)$ be the time independent Hamiltonian of some quantum mechanical system in one dimension [3]. The Heisenberg equation of motion for the
position and momentum operators as functions of discrete time can be written as follows
\begin{align}
{i \over \varepsilon }\left( {q_{n+1}-q_n} \right)={1 \over 2}\left[ {q_{n+1}+q_n,H} \right]\\[1ex]
{i \over \varepsilon }\left( {p_{n+1}-p_n} \right)={1 \over 2}\left[ {p_{n+1}+p_n,H} \right]
\end{align}
where the position operator $q_n\equiv q(n\varepsilon )$, and momentum operator $p_n\equiv p(n\varepsilon )$ must satisfy $\left[ {q_n,p_n}
\right]=i,\forall n$.

Let $P_k(x)$ be an polynomial of the variable $x\equiv qp+pq$. It is easy to prove that
\begin{align*}
\left[ {q,P_k(x)} \right]=\left( {P_k(x+2i)-P_k(x)} \right)q\\[1ex]
\left[ {p,P_k(x)} \right]=\left( {P_k(x-2i)-P_k(x)} \right)p
\end{align*}
When $H$ is a polynomial of the type $P_k(x)$ we have 
\begin{equation*}
{i \over \varepsilon }\left( {q_{n+1}-q_n} \right)=\left[ {P_k(x+2i)-P_k(x)} \right]{1 \over 2}\left( {q_{n+1}+q_n} \right)
\end{equation*}
hence
\begin{equation}
q_{n+1}={{1-{1 \over n}i\varepsilon \left( {P_k(x+2i)-P_k(x)} \right)} \over {1+{1 \over 2}i\varepsilon \left( {P_k(x+2i)-P_k(x)} \right)}}q_n
\end{equation}
By iteration we can calculate $q_n$ in terms of the initial condition $q_0$. Similarly
\begin{equation}
p_{n+1}={{1-{1 \over n}i\varepsilon \left( {P_k(x+2i)-P_k(x)} \right)} \over {1+{1 \over 2}i\varepsilon \left( {P_k(x+2i)-P_k(x)} \right)}}p_n
\end{equation}
Since $P_k(x)$ is hermitian we have $\left[ {q_{n+1},p_{n+1}} \right]=\left[ {q_n,p_n} \right]$. 

When $H$ is a function of $x$ we can expand it in
terms of some orthornomal polynomials of the variable $x$. In particular if we take the continuous Hahn polynomials $S_k(x)$ defined by the two terms
recursion relation
\begin{equation*}
kS_k(x)=xS_{k-1}(x)-(k-1)S_{k-2}(x)
\end{equation*}
we can express the totally symmetrie polynomial $T_{k,k}(q,p)$ of all posible monomials containing $k$ factors of $q$ and $k$ factors of $p$ as [10]
\begin{equation*}
T_{k,k}(q,p)={{(2k)!} \over {k!2^k}}S_k(qp+pq)
\end{equation*}
In all these cases we have
\begin{align}
q_n=\left[ {{{1-{1 \over 2}i\varepsilon \left( {H(x+2i)-H(x)} \right)} \over {1+{1 \over 2}i\varepsilon \left( {H(x+2i)-H(x)} \right)}}}
\right]^nq_0 \\[1ex]
 p_n=\left[ {{{1-{1 \over 2}i\varepsilon \left( {H(x+2i)-H(x)} \right)} \over {1+{1 \over 2}i\varepsilon \left(
{H(x+2i)-H(x)} \right)}}} \right]^np_0
\end{align}
which in the limit $n\longrightarrow \infty $, $\varepsilon\longrightarrow 0 $, $n\varepsilon\longrightarrow t $, become
\begin{align*}
q_n\longrightarrow q(t)=\exp \left( {itH(x)} \right)q(0)\exp \left( {-itH(x)} \right)\\[1ex]
p_n\longrightarrow p(t)=\exp \left( {itH(x)} \right)p(0)\exp \left( {-itH(x)} \right)
\end{align*}

\section{Dirac equation on the lattice}

Given a function $\psi (n_\mu )$ defined on the grid points of a Minkowski lattice with elementary lengths $\varepsilon_{\mu}$, difference operators $\Delta
(\nabla )$, average operators $\tilde \Delta (\tilde \nabla )$, we construct the Hamiltonian for the Dirac fields $\psi _\alpha (n_\mu )$ on the lattice [4]
\begin{align}
& H=\epsilon _1\epsilon _2\epsilon _3\sum\limits_{n_1n_2n_3=0}^{N-1} {\tilde \Delta _1}\tilde \Delta _2\tilde \Delta _3\;\psi ^+\left( {n_\mu}
\right) \times \nonumber \\
&  \times \left\{ {\gamma _0\gamma _1{1 \over {\varepsilon _1}}\Delta _1\tilde \Delta _2\tilde \Delta _3+\gamma _0\gamma _2\tilde \Delta _1{1 \over
{\varepsilon _2}}\Delta _2\tilde \Delta _3+\gamma _0\gamma _3\tilde \Delta _1\tilde \Delta _2{1 \over {\varepsilon _3}}\Delta _3+m_0c\gamma _0\tilde \Delta
_1\tilde \Delta _2\tilde \Delta _3} \right\}\psi (n_\mu )
\end{align}
from which we obtain (by the Hamilton equation of motion) the Dirac equation
\begin{equation}
\left( {i\gamma _\mu {1 \over {\varepsilon _\mu }}\Delta _\mu \prod\limits_{\nu \ne \mu } {\tilde \Delta _\nu -m_0c\;\tilde \Delta _0\tilde
\Delta _1\tilde
\Delta _2\tilde \Delta _3}} \right)\psi (n_\mu )=0
\end{equation}
the solution of which can be expressed in terms of the plane waves on the lattice, namely, the orthogonal
functions:
$$f\left( {n_\mu } \right)=\prod\limits_{\mu =0}^3 {\exp \left( {-i{{2\pi } \over N}m_\mu n_\mu } \right)\quad },\quad m_\mu,  n_\mu =0,1\ldots N-1$$
provided the dispersion relations are satisfied
\begin{equation}
k_\mu k^\mu =m_0^2c^2\quad ,\quad k_\mu \equiv {2 \over {\varepsilon _\mu }}tg{{\pi m_\mu } \over N}
\end{equation}
The transfer matrix, which carries the Dirac field from one time to the next time step can be obtain from the evolution operator
\begin{equation}
U={{1+{1 \over 2}\,i\,\varepsilon _0\,H} \over {1-{1 \over 2}\,i\,\varepsilon _0\,H}}\quad ,\quad \psi \left( {n_0+1} \right)=U\psi \left( {n_0} \right)U^+
\end{equation}
Our model for the fermion field on the lattice satisfies the following conditions in order to escape the no-go theorem of Nielsen-Ninomiya.

\begin{enumerate}
\item[i)] the Hamiltonian is traslational invariant

\item[ii)] the Hamiltonian is Hermitian

\item[iii)] for $m_0=0$, the wave equation is invariant under global chiral transformation

\item[iv)] there is no fermion doubling

\item[v)] the Hamiltonian is non-local (its Fourier transform has a singularity in the Brillouin zone) but the evolution operator, due to the Stone theorem, is
unitary.
\end{enumerate}

\bigskip{\noindent \Large \bf Acknoledgments}

\medskip
This work has been partially supported by Ministerio de Ciencia y Tecnolog\'{\i}a (Spain) under grant BFM2000-0357.

\vspace{2cm}

e-mail: mlp@pinon.ccu.uniovi.es


\begin{thebibliography}{00}
\bibitem{} J.C. Baez, An Introduction to Spin Foam Model of Background free Theory and Quantum Gravity, in: (H.
Gausteter, et al. eds.) {\em Geometry and Quantum Physics},  Lect. Not. Phys 543, (Springer 2000).
\bibitem{} D. Levy, L. Vinet, P. Winternitz (ed.) {\em Symmetries and integrability of Difference Equations}, (American Mathematical Society, 1996).
\bibitem{} M. Lorente, On some integrable one-dimensional quantum systems, {\em Phys. Lett, B 232} (1989) 345-350.
\bibitem{} M. Lorente, A new scheme for the Klein-Gordon and Dirac fields on the lattice with axial anomaly, {\em J. Group Theor. Phys. 1} (1993) 105-121.
\bibitem{} M. Lorente, Continuous vs discrete models for the Quantum Harmonic oscillator and the Hydrogen Atom, {\em Phys. Lett. A 285} (2001)
119-126.
\bibitem{} M. Lorente, Raising and lowering operators, factorization and differential/difference operators of hypergeometric type, {\em J. Phys. A:
Math. Gen. 34} (2001) 569-588.
\bibitem{} I. Montvay, G. M\"unster, {\em Quantum Field Theories on a Lattice}, (Cambridge U. Press 1994).
\bibitem{} A.F. Nikiforov, S.K. Suslov, V.B. Uvarov, {\em Classical Orthogonal polynomials of Discrete Variable}, (Springer 1991).
\bibitem{} A.M. Perelomov, Quantum integrable systems and Clebsch-Gordon series, {\em J. Phys. A: Math. Gen. 31} (1998) L31-L37.
\bibitem{} This formula was proved rigurously by T.H. Koornwinder, ``Meixner-Pollaczek polynomials and the Heisenberg algebra'', {\em J. Math. Phys. 30} (1989)
767-769.
\end{thebibliography}
\end{document}